\documentclass
[superscriptaddress,secnumarabic,amssymb,amsmath,nobibnotes,aps,prd,showkeys,showpacs,nofootinbib,onecolumn,notitlepage,10pt]{revtex4}%
\usepackage{setspace}
\usepackage{color}
\usepackage{amsmath}
\usepackage{amsfonts}
\usepackage{verbatim}
\usepackage{amssymb}
\usepackage{graphicx,bm}
\usepackage{amsmath}
\usepackage{amssymb}
\usepackage{graphicx}
\usepackage{subcaption}
\usepackage[caption=false]{subfig}%
\providecommand{\U}[1]{\protect\rule{.1in}{.1in}}
\newcommand{\be}{\begin{equation}}
\newcommand{\ee}{\end{equation}}

\newcommand{\mincir}{\raise
-3.truept\hbox{\rlap{\hbox{$\sim$}}\raise4.truept\hbox{$<$}\ }}
\newcommand{\magcir}{\raise
-3.truept\hbox{\rlap{\hbox{$\sim$}}\raise4.truept\hbox{$>$}\ }}

\ifx\pdfoutput\relax\let\pdfoutput=\undefined\fi
\newcount\msipdfoutput
\ifx\pdfoutput\undefined\else
\ifcase\pdfoutput\else
\ifx\paperwidth\undefined\else
\ifdim\paperheight=0pt\relax\else\pdfpageheight\paperheight\fi
\ifdim\paperwidth=0pt\relax\else\pdfpagewidth\paperwidth\fi
\fi\fi\fi
\begin{document}
\title{The \textquotedblleft Telephone Game\textquotedblright\ Effect in Modern
Gravity Research}
\author{Andronikos Paliathanasis}
\email{anpaliat@phys.uoa.gr}
\affiliation{School for Data Science and Computational Thinking and Department of
Mathematical Sciences, Stellenbosch University, Stellenbosch, 7602, South Africa}
\affiliation{Centre for Space Research, North-West University, Potchefstroom 2520, South Africa}
\affiliation{Departamento de Matem\`{a}ticas, Universidad Cat\`{o}lica del Norte, Avda.
Angamos 0610, Casilla 1280 Antofagasta, Chile}
\affiliation{National Institute for Theoretical and Computational Sciences (NITheCS), South Africa}

\begin{abstract}
Triggered by some recent studies within the framework of nonmetricity gravity,
I discuss how the basic information regarding the equivalence of the
gravitational theories forming the \textquotedblleft Trinity of
Gravity\textquotedblright\ has been lost in parts of the literature. As a
result, several studies with focus in modified theories of gravity have
research investigated the case of STEGR/GR, often without acknowledging this
fundamental equivalence or recognizing that their results can be derived in GR

\end{abstract}
\maketitle

Modified theories of gravity have drawn the attention of the society because
they can provide a geometric framework for the description of gravitational
phenomena. From a physical perspective, they are of particular interest as
they help us explore the role of geometric invariants in gravity. On the other
hand, from a mathematical point of view, these theories introduce new degrees
of freedom and lead to gravitational systems with rich and novel properties.
This has led to a wide range of studies in the literature.

The Ricci scalar $R$, the torsion scalar $T$ for a antisymmetric connection,
and the nonmetricity $Q$ are constructed by the three different components of
a generic connection $\Gamma_{\mu\nu}^{\kappa}$, the Levi-Civita part, the
antisymmetric part and the nonmetricity component respectively \cite{eis}.
These scalars are known as the trinity of gravity \cite{trin}, because the
variation of the Action Integral%
\begin{equation}
S=\int d^{4}x\sqrt{-g}\left(  \mathcal{A}+const\right)  ,~\mathcal{A=}\left(
R,T,Q\right)
\end{equation}
lead to the the same field equations.

General Relativity (GR) is the case where $\,\mathcal{A}$ is replaced by the
Ricci scalar $R$, when $T$ is used we end with the Teleparallel Equivalent
General Relativity (TEGR)~\cite{tegr}, while when $\mathcal{A}$ is replaced by
the nonmetricity scalar $Q$, we end with the Symmetric Teleparallel Equivalent
General\ Relativity (STEGR)~\cite{stegr}.

The elements of the trinity of gravity are connected with the following
relations \cite{trin}%
\begin{align*}
T  &  =R+B_{T},\\
Q  &  =R+B_{Q},
\end{align*}
where $B_{T}$ and $B_{Q}$ are topological boundary terms.

It is well known from the analytic mechanics that the introduction of a
boundary term in the Action Integral/Lagrangian does not affect the equations
of motion which follow from the variational principle, that is, the
Euler-Lagrange equations. Thus, that equivalency is lost in $f\left(
\mathcal{A}\right)  $ theories with $f$ to be a nonlinear function. This basic
property of these three geometric scalars has often been neglected in the
literature, and there are various studies in the literature where the authors
start from a modified theory of gravity but they end to study the case of GR.
As I will describe in the following lines this important information has been
got distorted in a part of the recent literature as it was passed along the
last few years. The lost of information reminds me the "Telephone Game" effect.

Recently, in \cite{epjco} the authors claimed that they investigated a static
spherical symmetric spacetime within the $f\left(  Q\right)  $. However, they
assumed that $f\left(  Q\right)  $ is linear, that is, $f\left(  Q\right)
=\beta_{0}Q+\beta_{1}$. Consequently, they reproduced previous results of
STEGR/GR. But that is not the only example. 

In \cite{pp1} the authors claimed that they study the characteristics of
spherically symmetric anisotropic compact objects within the framework of
$f\left(  Q\right)  $-gravity. Again they consider a linear function $f\left(
Q\right)  $, which means that the gravitational model is again that of
STEGR/GR, and they recovered the analysis presented decades ago in
\cite{grr}.\ Only the last weeks a linear function $f\left(  Q\right)  $ was
introduced in various of studies of $f\left(  Q\right)  $-gravity
\cite{pp6,pp7,pp8,pp2}. Some older studies in $f\left(  Q\right)  $ theory
that considered a linear function $f\left(  Q\right)  $, are the following,
though this list is not exhaustive \cite{kp1,kp2,kp3,kp4,kp5}.

Within the framework of $f\left(  Q,B_{Q}\right)  $ theory of gravity, there
are many studies where the authors consider linear dependence on $B_{Q}$ which
lead to the $f\left(  Q\right)  -$gravity, or linear dependence on both of the
scalars, which lead to STEGR, see for instance \cite{qq1,qq2,qq3}. \ 

I believe that my point is clear and that it is not any need to extend the
discussion further, mention more works with the same consideration nor to
extend the discussion to other theories of gravity where $T,$ $Q$, $B_{T}$,
$B_{Q}~$or other boundary terms, appear as linear components in the
gravitational Action Integral. However, it is important to note that some of
the journals publishing the aforementioned studies have been informed of this
remark, but not all have shown interest in addressing it.

It is clear that readers should be careful and think critically when
addressing these studies. Last but not least, reviewers should be even more
careful and thoughtful, as they carry part of the responsibility for the
publication of this kind of works, in order the \textquotedblleft Telephone
Game\textquotedblright\ effect to be ended.

\bigskip

\textbf{Data Availability Statement:} Data sharing is not applicable to this
article as no datasets were generated or analyzed during the current study.

\bigskip

\textbf{Code Availability Statements:} Code sharing is not applicable to this
article as no code generated by the current study.

\begin{acknowledgments}
The author thanks the support of Vicerrector\'{\i}a de Investigaci\'{o}n y
Desarrollo Tecnol\'{o}gico (Vridt) at Universidad Cat\'{o}lica del Norte
through N\'{u}cleo de Investigaci\'{o}n Geometr\'{\i}a Diferencial y
Aplicaciones, Resoluci\'{o}n Vridt No - 096/2022 and Resoluci\'{o}n Vridt No -
098/2022. The authors was partially supported by Proyecto Fondecyt Regular
2024, Folio 1240514, Etapa 2024.
\end{acknowledgments}

\end{document}